\documentclass[10pt,aps,prl,showpacs,superscriptaddress,twocolumn,citeautoscript]{revtex4-1}

\usepackage{graphicx}  % needed for figures
\usepackage{dcolumn}   % needed for some tables
\usepackage{bm}        % for math
\usepackage{amssymb}  
\usepackage{amsmath}
\usepackage{hyperref}

\newcommand{\beq}{\begin{equation}}
\newcommand{\eeq}{\end{equation}}

\newcommand{\subeqs}[2]{\begin{subequations}\label{#2}\begin{align} #1 \end{align}\end{subequations}}

\begin{document}
\title{Non-Reciprocal Thermal Material by Spatio-Temporal Modulation}
\author{Daniel Torrent}
\email{torrent@crpp-bordeaux.cnrs.fr}
\affiliation{Centre de Recherche Paul Pascal, UPR CNRS 8641, Universit\'e de Bordeaux, Pessac, (France)}
\affiliation{GROC, UJI, Institut de Noves Tecnologies de la Imatge (INIT), Universitat Jaume I, 12080, Castell\'o, (Spain)}
\author{Olivier Poncelet}
\author{Jean-Chirstophe Batsale}
\affiliation{Institut de M\'ecanique et d'Ing\'enierie, UMR CNRS 5295, Universit\'e de Bordeaux, Talence (France)}
\date{\today}
%%%%%%%%%%%%%%%%%%%%%%%%%%%%%%%%%%%%%%%%%%%%%%%%%%%%%%%%%%%%%%%%%%%%%%%%%%%%%%%%%%%%%%%%%%%

%%%%%%%%%%%%%%%%%%%%%%%%%%%%%%%%%%%%%%%%%%%%%%%%%%%%%%%%%%%%%%%%%%%%%%%%%%%%%%%%%%%%%%%%%%%
\begin{abstract}
The thermal properties of a material with a spatio-temporal modulation in both the thermal conductivity and the mass density are studied. The special configuration studied here consists of a modulation in a wave-like fashion. It is found that these materials behaves, in an effective way, as materials with an internal convection-like term that provides them of non-reciprocal properties, in the sense that the flow of heat has different properties when it propagates in the same direction or in the opposite one to the modulation of the parameters. An effective medium description is presented which accurately describes the modulated material, and numerical simulations supports both the non-reciprocal properties and the effective medium description. It is found that these materials are promising candidates for the design of thermal diodes and other advanced devices for the control of the heat flow at all scales.

\end{abstract}
%%%%%%%%%%%%%%%%%%%%%%%%%%%%%%%%%%%%%%%%%%%%%%%%%%%%%%%%%%%%%%%%%%%%%%%%%%%%%%%%%%%%%%%%%%%
\maketitle
%%%%%%%%%%%%%%%%%%%%%%%%%%%%%%%%%%%%%%%%%%%%%%%%%%%%%%%%%%%%%%%%%%%%%%%%%%%%%%%%%%%%%%%%%%%

The research on materials with non-reciprocal thermal properties has received a great attention in recent years. These materials have different propagation properties of the thermal energy along two opposite directions. With the so-called thermal diode being the most immediate application of these structures\cite{li2004thermal}, other devices and applications are easily envisioned, like thermal transistors and even logic circuits\cite{wang2007thermal}.

Non-reciprocal materials have been properly studied theoretical and experimentally at different scales \cite{,liang2009acoustic,kobayashi2009oxide,budaev2016thermal,romero2016thermal}, and it has been demonstrated that the realization of a non-reciprocal material requires in general the use of a combination of non-linear and asymmetric structures\cite{roberts2011review}. However, the realization of non-reciprocal materials based on non-linear elements limits their applicability, since non-linearity does not occurs at all temperatures and scales, so that we can find that the rectification properties of the materials be efficient in only a short temperature range. 

In this context, phononic metamaterials, artificially structured materials with {\it a priori}-designed properties, have overcome one of the major drawbacks of common materials, since their properties depend on the internal artificial structure and not on intrinsic properties of the constituent materials, which in turns allow us to decide at which scale and frequency or temperature range we want to operate\cite{maldovan2013sound}. Therefore, a special type of metamaterial is employed in this work presenting non-reciprocal properties. 

The proposed metamaterial consist in a periodically modulated thermal material, however this modulation will happen in both space and time. This special type of modulation has been studied in elastic and acoustic materials\cite{swinteck2015bulk,trainiti2016non,nassar2017modulated,nassar2017non}, whose non-reciprocal properties for the propagation of waves have been widely demonstrated. In this work we will apply these ideas to the diffusion equation describing thermal waves in solids, and it will be found that the behavior of this equation is completely different to the classical wave equations.

We present therefore an alternative mechanism for the realization of non-reciprocal thermal materials which, in principle, can be applied to any scale. The mechanism consist in the realization of materials in which the thermal properties (conductivity and mass density) are modulated in both space and time, so that it is demonstrated that, when the spatio-temporal modulation of these properties are of the form of a traveling wave, the material presents non-reciprocal thermal properties, and the heat flow is allowed only in the direction of the traveling modulation. Moreover, it is demonstrated that an effective medium description is possible for such a material, in which it is described as a homogeneous solid with constant constitutive parameters (in both space and time) but in which the temperature field satisfies the convection-diffusion equation. In other words, it is demonstrated that, although there is no transport of matter in the solid material, in an effective way an internal convective term appear, which is the responsible of providing non-reciprocal properties to the solid even in the stationary regime. Analytical expressions are given for the effective parameters and time-domain numerical simulations performed by the commercial software COMSOL show a perfect agreement with the effective medium description.

\begin{figure}[ht!]
\begin{center}
\includegraphics[width=\columnwidth]{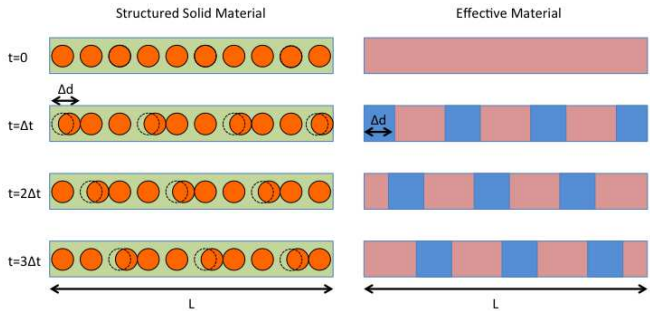}%
\caption{\label{fig:stmaterial}Schematic representation of a possible realization of a material with a spatio-temporal modulation in the conductivity and the mass density.}%
\end{center}
\end{figure}

Figure \ref{fig:stmaterial} shows a possible realization (though not the only one) of a material with a spatio-temporal modulation in both the mass density and conductivity. In the left panel, upper part, a linear chain of spheres of length $L$ is surrounded by a solid background. This system behaves as an effective material with some effective mass density and conductivity, as represented in the right panel. It is obvious that the effective properties of this material will depend, among other parameters, on the distance between spheres. Therefore, if at $t=\Delta t$ a periodic perturbation is introduced in the material so that the distance between spheres is changed a quantity $\Delta d$ every three spheres, the effective parameters of the chain of spheres will be different in this region, and a periodic modulation will have been induced, as it is shown in the right panel. Finally, if this perturbation is traveling through the material, the effective material will have also a traveling-like behavior, as it is shown in the last two panels of figure \ref{fig:stmaterial}, where the perturbation travels at a speed $v_0\approx d/\Delta t$.
%%%%%%%%%%%%%%%%%%%%%%%%%%%
%%%%%%%%%%%%%%%%%%%%%%%%%%%

The mechanism described before and represented in figure \ref{fig:stmaterial} is obviously not unique, and it is just an example of how a spatio-temporal modulation in the thermal properties of a solid can be induced by means of a perturbation (or an external field). The objective in this work is not to study the possible realization of this special modulation, but to study the properties of a material in which we can assume that the thermal conductivity $\sigma$ and mass density $\rho$ are of the form
\subeqs{
\sigma&=\sigma(x-v_0t),\label{eq:sigma}\\
\rho&=\rho(x-v_0t),\label{eq:rho}
}{eq:params}
with $\sigma$ and $\rho$ being periodic functions of $n=x-v_0t$ with period $d$. In a material with these properties, the energy balance is described by means of the diffusion equation
\beq
\label{eq:diffxt}
\frac{\partial}{\partial x}\left(\sigma(x-v_0t)\frac{\partial T}{\partial x}\right)=\rho(x-v_0t)\frac{\partial T}{\partial t},
\eeq
where we have stated that the heat capacity is equal to 1, in order to simplify the notation, however it is evident that in the above equation $\rho$ means the product $\rho c_V$.

We are interested in the regime in which the spatio-temporal variation of the constitutive parameters is not ``visible'' and we perceive the material as a homogeneous material. This is the classical homogenization limit, in which heterogeneous materials are perceived as effective materials with some averaged properties, therefore we want to know here which ones are the effective mass density and conductivity of a material with this special modulation in the materials' parameters. It must be pointed out that the effective material description is valid under situations in which the variations of the fields (the temperature here) are smoother than the micro-structure variations. In the following lines it will be shown that the homogeneous version of equation \eqref{eq:diffxt} contains additional constitutive parameters that provides of this material of non-reciprocity.

The homogenization of equation \eqref{eq:diffxt} can be done more efficiently under the change of variables $n=x-v_0t$ and $\tau=t$, so that the diffusion equation takes the form
\beq
\label{eq:Tnframe}
\frac{\partial}{\partial n}\left(\sigma(n)\frac{\partial T}{\partial n}\right)=\rho(n)\frac{\partial T}{\partial \tau}-\rho(n) v_0 \frac{\partial T}{\partial n},
\eeq
which is an equation in which the coefficients depend only on the variable $n$. The transformed equation in the $n-\tau$ coordinate system (actually a traveling  reference frame at velocity $v_0$) is the diffusion-convection equation with periodic coefficients in the coordinate $n$, where the coefficient $\rho(n)v_0$ appears as a convection coefficient representing a flow of matter in the opposite direction of the traveling modulation. This is somehow obvious since the traveling frame sees the material traveling in the opposite direction, since the material is actually at rest.

Equation \ref{eq:Tnframe} is a partial differential equation in the variables $n$ and $\tau$ in which the coefficients are periodic functions of $n$ with period $d$, so that Block theorem applies and the solution for the temperature field is a linear combination of eigenfunctions of the form
\beq
\label{eq:Tblock}
T(n,\tau)= e^{-iKn}e^{i\Omega \tau} \phi(n), 
\eeq
with $\phi(n)$ being a periodic function of the variable $n$ with the same periodicity of $\sigma$ and $\rho$. When this form of the temperature field is introduced in equation\eqref{eq:Tnframe} the dispersion relation $\Omega=\Omega(K)$ is finally obtained as the solution of an eigenvalue problem.

The spatio-temporal behavior of the temperature field is therefore composed of the ``macroscopic'' function $e^{-iKn}e^{i\Omega \tau}$ modulated by a ``microscopic'' function $\phi(n)$. When the spatial variations of the field are larger than the typical period $d$, we are in the so called homogenization limit, and equation \eqref{eq:Tnframe} can be replaced by a ``homogenized'' version with constant coefficients with the same solution $\Omega=\Omega(K)$. Once the equation in the traveling frame is homogenized, we can return to the frame at rest to study its properties, as described in the Supplementary Material. However, when we return to the system at rest, we don't recover equation \eqref{eq:diffxt} with constant coefficients, as should be expected from a homogenization process, but we obtain a more complicated equation, in which additional constitutive parameters appear,
\beq
\label{eq:diffconvection}
\sigma^*\frac{\partial^2 \langle T\rangle }{\partial x^2}= \rho^*\frac{\partial\langle T\rangle }{\partial t}+C\frac{\partial\langle T\rangle }{\partial x}-i(S+S')\frac{\partial^2 \langle T\rangle }{\partial x\partial t}.
\eeq
%%%%%%%%%%%%%%%%%%%%%%%%%%%
\begin{figure}[ht!]
\begin{center}
\includegraphics[width=\columnwidth]{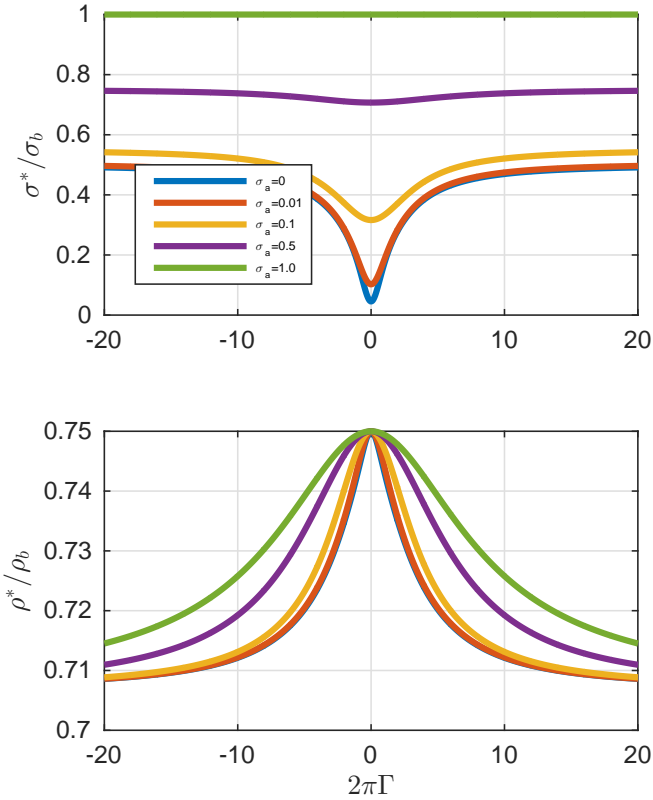}%
\caption{Effective thermal conductivity (upper panel) and mass density (lower panel) as a function of the non-dimensional modulation velocity $\Gamma$.}%
\label{fig:sigmarho}
\end{center}
\end{figure}
%%%%%%%%%%%%%%%%%%%%%%%%%%%

Therefore, the homogenized equation is the convection-diffusion equation with two additional coefficients, $S$ and $S'$, which are the thermal equivalent of the Willis coefficients found in the homogenization of phononic crystals\cite{norris2012analytical}. These coefficients are coupling terms related with the non-symmetry of the unit cell, and although they are null for symmetric periodic materials\cite{torrent2015resonant}, the non-reciprocity induced by the special modulation of the materials considered here makes them different than zero. These terms are relevant specially in the dynamic or transient regime, however in this work we are more interested in the non-reciprocal properties of the material in the nearly stationary regime, for which a further discussion about these terms is beyond the objective of the present work.
%%%%%%%%%%%%%%%%%%%%%%%%%%%
\begin{figure}[ht!]
\begin{center}
\includegraphics[width=\columnwidth]{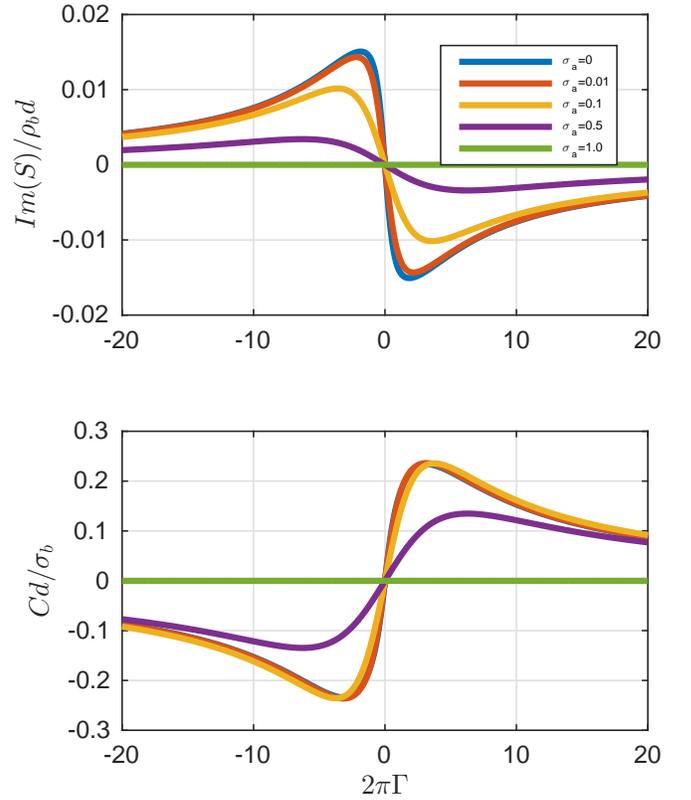}%
\caption{Effective Wills term (upper panel) and convection coefficient (lower panel) as a function of the non-dimensional modulation velocity $\Gamma$.}%
\label{fig:SC}
\end{center}
\end{figure}
%%%%%%%%%%%%%%%%%%%%%%%%%%%

The responsible of the non-reciprocal properties of the material in the stationary regime is the convective term $C\partial_x T$ appearing in equation \eqref{eq:diffconvection}. It is interesting the relationship between the convective term $C$ and the effective mass density $\rho^*$. It could be thought that, since $v_0$ is constant through the material, the effective convective term in the homogenized version of equation \eqref{eq:Tnframe} would be simply $v_0\rho^*$. The consequence of this property would be that, when returning to the rest reference frame, the convective term would disappear and then we would recover the diffusion equation with constant coefficients (plus the Willis terms). However, as it is demonstrated in the Supplementary Material, the effective convective term does not satisfy this condition, since although the variation of $v_0\rho$ is the same as of $\rho$, they appear multiplying a different operator in the equation, the temporal derivative and the spatial derivate, so that their role is completely different in the equation and, therefore, in the frame at rest we find that the diffusion equation \eqref{eq:diffxt} has become the diffusion-convection equation \eqref{eq:diffconvection}, which is known to be non-reciprocal due to the convective term $C$.

Therefore, the spatio-temporally modulated material behaves, in the homogenization limit, as a homogeneous material in which a convective term appear, so that the propagation of heat will have non-reciprocal properties. It must be pointed out that the convective term is not induced by any transport of matter, as in fluid dynamics and similar processes, but it is induced by means of some external stimulus which is modulating the properties of the material in a wave-like fashion, so that we can have not only a solid material with an internal effective convection, but we can have a finite structure with convection without the need of letting the flow of matter leave the structure. 

The Supplementary Material shows the mathematical expressions to obtain these effective parameters for an arbitrary form of the periodic functions $\sigma(x-v_0t)$ and $\rho(x-v_0t)$. The simpler case of modulation is a simple cosinus perturbation of the form
\subeqs{
\sigma(x-v_0t)&=\sigma_0\left[1+\Delta_\sigma\cos\frac{2\pi}{d}(x-v_0t)\right],\\
\rho(x-v_0t)&=\rho_0\left[1+\Delta_\rho\cos\frac{2\pi}{d}(x-v_0t)\right],
}{eq:harmonic}
where the mass density and conductivity changes periodically from $\rho_b=\rho_0(1-\Delta_\rho)$ to $\rho_a=\rho_0(1+\Delta_\rho)$ and from $\sigma_b=\sigma_0(1-\Delta_\sigma)$ to $\sigma_a=\sigma_0(1+\Delta_\sigma)$, respectively. The effective parameters in this situation can be approximated by (see equations 32 in the Supplementary Material)
\subeqs{
\sigma^*&\approx\sigma_0\left[1-\frac{1}{2}\frac{\Delta_\sigma^2}{1+\Gamma^2}\right],\\
\rho^*&\approx\rho_0\left[1-\frac{\Gamma^2}{2}\frac{ \Delta_\rho^2}{1+\Gamma^2}\right],\\
S=S'&\approx-\frac{\rho_0d}{2\pi}\frac{\Delta_\rho\Delta_\sigma}{2}\frac{i\Gamma}{1+\Gamma^2},\\
C&\approx\frac{2\pi\sigma_0}{d} \frac{\Delta_\rho\Delta_\sigma}{2}\frac{\Gamma}{1+\Gamma^2},
}{eq:hparams}
where we have defined the normalized modulation velocity $\Gamma$ as
\beq
\Gamma=\frac{v_0 d \rho_0}{2\pi \sigma_0}.
\eeq

Equations \eqref{eq:hparams} show that the effective conductivity and mass density are both even functions of $\Gamma$, meaning that reversing the direction of the modulation has no effect on their values. Contrarily, both $S$ and $C$ are odd functions, which is obvious since these parameters are the responsible of the non-reciprocal properties of the material. When there is no traveling modulation ($\Gamma=0$), both $S$ and $C$ are zero, the mass density is just the average mass density $\rho^*=\rho_0$ and effective conductivity $\sigma^*=\sigma_0(1-\Delta_\sigma^2/2)$, so that we recover reciprocity as expected. Interestingly, when $v_0\to\pm\infty$ the non-reciprocal properties of the material also disappear, since $S$ and $C$ both tend to zero, and now the effective mass density is $\rho^*=\rho_0(1-\Delta_\rho^2/2)$ and the effective conductivity is $\sigma^*=\sigma_0$. In this case the oscillations of the material's properties are so fast that the spatial variation almost disappear, therefore we can see an averaged material in time, which in turns means that the non-reciprocal properties disappears. It is interesting to note how the expressions for the effective parameters exchange their roles in the limiting situation $\Gamma=\pm \infty$ or $\Gamma=0$, due to the exchange of them in front of the space and time derivatives in the diffusion equation. This simple analysis, which will be verified later, shows that the larger ``non-reciprocity'' is not obtained increasing the modulation velocity, but that there is an optimum velocity for the design of non-reciprocal materials.

Another interesting feature of equations \eqref{eq:hparams} is that we need a modulation of both the mass density and the thermal conductivity to have non-reciprocity. This is indeed a general result, as shown in the Supplementary Material, where the effective convective term is shown to be
\beq
C=v_0\sum_{G',G\neq 0}\rho_{-G'}G'\chi_{G'G}\sigma_{G}G
\eeq
where the summation has to be performed for all the reciprocal lattice points $G=2\pi m/d$, with $m$ being an integer. $\chi_{G'G}$ is an interaction matrix, and $\rho_G$ and $\sigma_G$ are the Fourier components of the functions $\rho(n)$ and $\sigma(n)$, respectively. Given that in the above equation the summation excludes the term $G=0$, it will be zero unless we have at least one pair ($\rho_G, \sigma_G$) for $G\neq 0$ different than zero, that is, we need a simultaneous variation of both $\sigma$ and $\rho$.

 This result shows that the origin of the convective term in the effective material is due to a coupling between the variation of the mass density and the conductivity, and enforces its analogy with the Willis term and chirality in electromagnetism. 

Figure \ref{fig:sigmarho} shows the dependence on $\Gamma$ of the effective conductivity (upper panel) and mass density (lower panel) relative to those of the background ($\sigma_b$ and $\rho_b$, respectively) computed by means of the expressions 27 of the Supplementary Material. In these examples $\rho_a/\rho_b=0.5$ and $\sigma_a/\sigma_b=0,0.01,0.1,0.5$ and 1, as indicated in the legends of the plots. The saturation effect as $\Gamma\to\infty$ is evident.

Figure \ref{fig:SC} shows the dependence on $\Gamma$ of the parameters responsible of the non-reciprocal properties of the effective material, $S$ and $C$. As discussed before, these parameters change of sign when $\Gamma$ does the same. Also, besides the effect of saturation discussed before, which cancels non-reciprocity for $\Gamma\to\pm \infty$, we see that there is a value of $\Gamma=\Gamma_M$ for which both $S$ and $C$ have a local maximum, which defines the optimum $v_0$  to achieve non-reciprocity.

In the stationary regime the macroscopic temperature $\langle T\rangle$ is independent of time, and equation \eqref{eq:diffconvection} reduces to
\beq
\sigma^*\frac{\partial^2 \langle T\rangle }{\partial x^2}= C\frac{\partial\langle T\rangle }{\partial x}
\eeq
whose solutions are given by
\beq
\label{eq:Tx}
\langle T\rangle =A+Be^{\alpha x},
\eeq
with $\alpha=C/\sigma^*$. The above equation shows clearly the non-reciprocal nature of the distribution of temperature, as well as of the heat flux $\Phi(x)=\sigma^*\partial_xT(x)$. The larger the parameter $\alpha$, the higher the non-reciprocity of the material. For the harmonic perturbation studied in the present example, we can approximate $\alpha$ by
\beq
\alpha\approx\frac{2\pi}{d}\Delta_\sigma\Delta_\rho\frac{\Gamma}{1+2\Gamma^2}.
\eeq

\begin{figure}[ht!]
\begin{center}
\includegraphics[width=\columnwidth]{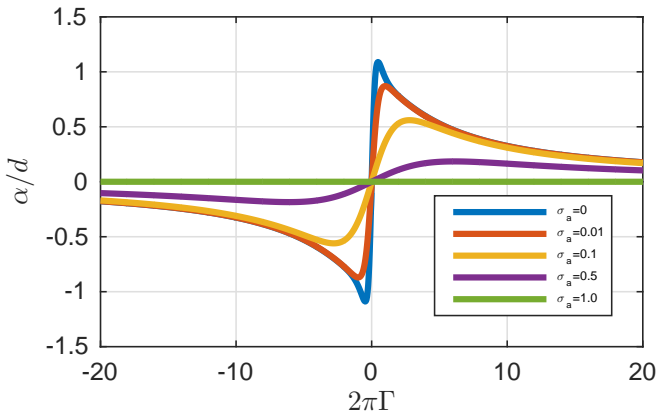}%
\caption{\label{fig:alpha} Effective convection-diffusion coefficient as a function of the non-dimensional modulation velocity $\Gamma$.}
\end{center}
\end{figure}

Figure \ref{fig:alpha} shows the dependence of this parameter as a function $2\pi\Gamma$, we see that there is an optimum value of $\Gamma$ for which we obtain the maximum value of $\alpha$ and, as before for $C$, when $\Gamma\to\infty$, $\alpha$ tends to zero and the material becomes reciprocal.
%%%%%%%%%%%%%%%%%%%%%%%%%%%
%%%%%%%%%%%%%%%%%%%%%%%%%%%

In order to check the accuracy of this description, we have performed numerical simulations in time-domain with the commercial software COMSOL multiphysics. In these simulations, we have assumed a one dimensional domain (a solid bar, for instance) of length $L=10d$, in which the initial temperature is set to 0. In the ``forwards'' (F) configuration, the temperature at the extreme $x=L$ is fixed to 0 and, for $t>0$, the temperature at $x=0$ is set to $T_0$. In the ``backwards'' configuration we have reversed the temperatures, so that at $x=0$ the temperature is fixed to 0 and for $t>0$ the temperature is fixed to $T_0$ at $x=L$. We have selected the same parameters for $\rho_a$ and $\rho_b$ as in the previous calculations, and the value of $\sigma_a=0.01\sigma_b$. The simulations have been performed for $2\pi\Gamma=0,0.3,1$ and $10$, whose corresponding values for $\alpha$ are shown as red points in figure \ref{fig:alpha}. According to equation \eqref{eq:Tx} and the previously defined boundary conditions, the temperature distribution in the bar in the stationary regime for the forwards and backwards configuration is, respectively,
\subeqs{
\langle T_F\rangle&=T_0\frac{e^{\alpha L}-e^{\alpha x}}{e^{\alpha L}-1}\\
\langle T_B\rangle&=T_0\frac{e^{\alpha x}-1}{e^{\alpha L}-1}
}{eq:Tbar}

\begin{figure}[ht!]
\begin{center}
\includegraphics[width=\columnwidth]{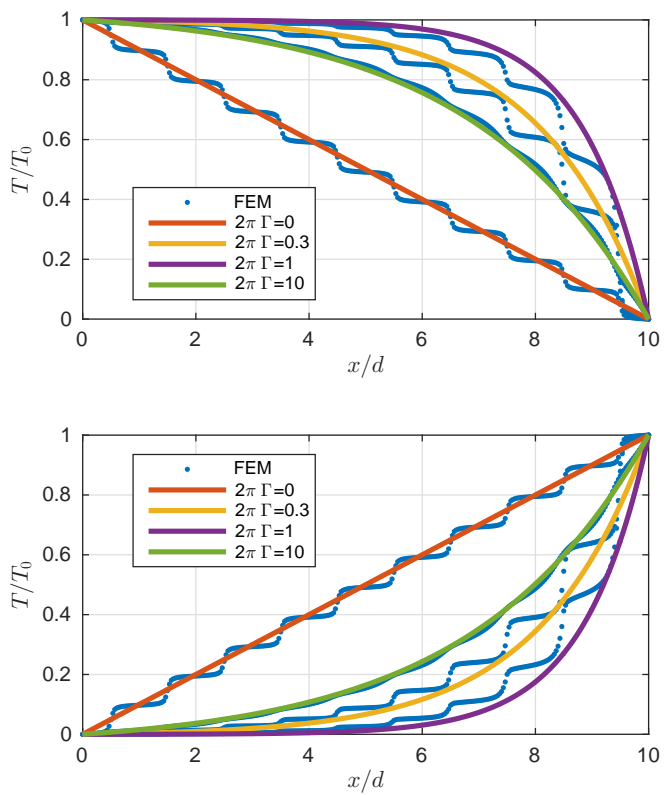}%
\caption{\label{fig:Tx} Temperature distribution of the spatio-temporally modulated bar in the forward (upper panel) and backward (lower panel) configurations.}
\end{center}
\end{figure}

Figure \ref{fig:Tx} shows the numerical simulations performed by COMSOL (blue dots) at $t=t_f=300 d\rho_b/\sigma_b$, together with the corresponding analytical solution given by \eqref{eq:Tbar}. It is clear that there is a perfect agreement with the numerical and analytical solution, although an additional modulation appears in the numerical simulation. This modulation is due to the fact that in the homogenized model we ignore the modulation function $\phi(n)=\phi(x-v_0t)$, which is obviously included in the numerical solution. Since the time is fixed to $t=t_f$ in figure \ref{fig:Tx}, only the spatial variation of $\phi$ is detected, however the transient period and the time evolution of the system can be seen in the Supplementary Movies temperatureF.gif and temperatureB.gif, where the effect of $\phi(n)$ is more evident. However, the relevant information is given by the analytical model shown in equation \eqref{eq:Tbar} with the $\alpha$ parameter computed by means of equations 27 in the supplementary material. It is obvious the diode-like behavior of the material, whose non-reciprocal nature is manifested not only in the static but also in the dynamic regime. The accuracy of the analytical solution provides also a very powerful tool to design more advanced devices based on these materials.
%%%%%%%%%%%%%%%%%%%%%%%%%%%
%%%%%%%%%%%%%%%%%%%%%%%%%%%

In summary, we have presented a structured solid material with non-reciprocal effective thermal properties, where the mechanism of non-reciprocity is due to an artificial convective term that appears in its effective behavior. The structured material consists of a modulated solid in which the thermal properties depend not only on the position, but also on time, in such a way that these parameters have a wave-like behavior. It is shown that, in the nearly-stationary regime and when the spatial variations of the sources are larger than the typical periodicity of the modulation, the material presents non-reciprocity in the propagation of heat, and it is shown how such a material can work as a thermal diode. Several properties of the effective parameters are deduced and an effective medium theory is developed. The expression derived for the convective term shows that it is required a modulation in both the mass density and thermal conductivity, since this term appears as a coupling between the relative variations of both parameters. Coupling terms equivalent to the so-called Willis terms in elasticity or chiral coefficients in electromagnetism also appear, although their contribution is relevant only in the transitory or time-dependent regime. It is remarkable the fact that the non-reciprocal thermal effect presented here is the result of the artificial internal structure of the materials, what makes that this effect be scalabe and therefore useful in a wide variety of thermal problems, going from the macro to nano scale.

\section{acknowledgements}
Work supported by the LabEx AMADEus (ANR-10- 444 LABX-42) in the framework of IdEx Bordeaux (ANR-10- 445IDEX-03-02), France. 

%\bibliographystyle{apsrev}
%\bibliography{bibliography}

\begin{widetext}
\section{Supplementary Material}
The diffusion equation for a material with parameters depending on both space and time in a wave-like fashion is given by
\beq
\label{eq:diffxt}
\frac{\partial}{\partial x}\left(\sigma(x-v_0t)\frac{\partial T}{\partial x}\right)=\rho(x-v_0t)\frac{\partial T}{\partial t}.
\eeq
The change of variable $n=x-v_0t$ and $\tau=t$ transform this equation in
\beq
\label{eq:diffntau}
\frac{\partial}{\partial n}\left(\sigma(n)\frac{\partial T}{\partial n}\right)=\rho(n)\frac{\partial T}{\partial \tau}-\rho(n) v_0 \frac{\partial T}{\partial n}
\eeq
where both $\sigma(n)$ and $\rho(n)$ are periodic functions of $n$. These functions can be expanded as Fourier series in the traditional way
\subeqs{
\sigma(n)&=\sum_G\sigma_Ge^{-iGn}\\
\rho(n)&=\sum_G\rho_Ge^{-iGn}
}{eq:FG}
and the solution for the temperature field can also be expressed in the form of a Bloch function
\beq
T(n,\tau)=e^{-ikn}e^{i\Omega \tau}\phi(n)=e^{-ikn}e^{i\Omega \tau}\sum_G\phi_Ge^{-iGn}
\eeq
where we have used the property of periodicity of $\phi(n)$. Inserting the above equations into equation \eqref{eq:diffntau} we arrive to the typical matrix equation defining the solutions $\Omega=\Omega(k)$,
\beq
-\sum_{G'}(k+G)\sigma_{G-G'}(k+G')T_{G'}=i\Omega\sum_{G'}\rho_{G-G'}T_{G'}+iv_0\sum_{G'}\rho_{G-G'}(k+G')T_{G'},
\eeq
we can now reorganize the second term of the right hand side of the above equation to group the term $\Omega+v_0k$,
\beq
-\sum_{G'}(k+G)\sigma_{G-G'}(k+G')T_{G'}=i(\Omega+v_0k)\sum_{G'}\rho_{G-G'}T_{G'}+iv_0\sum_{G'}\rho_{G-G'}G'T_{G'}.
\eeq
Notice that the term $\Omega+v_0k$ is actually the frequency in the $x-t$ frame, since the solutions of the equation in this frame are
\beq
T(n,\tau)=e^{-ikn}e^{i\Omega \tau}\phi(n)=e^{-ikx}e^{i(\Omega+v_0 k) t}\phi(x-v_0t)=e^{-ikx}e^{i\omega t}\phi(x-v_0t)
\eeq
therefore the replacement $k=k$ and $\omega=\Omega+v_0k$ returns the system of equations to the $x-t$ frame, giving
\beq
-\sum_{G'}(k+G)\sigma_{G-G'}(k+G')T_{G'}=i\omega\sum_{G'}\rho_{G-G'}T_{G'}+iv_0\sum_{G'}\rho_{G-G'}G'T_{G'}.
\eeq
The above equation will allow us to define the effective parameters of the material with the spatio-temporal modulation. We are interested now in the average temperature field, $T_0$, since it can be interpreted as the macroscopic temperature, therefore we split the above equation in two terms, those for $G=0$ and those for $G\neq 0$, 
\subeqs{
-k^2\sigma_0T_0-k\sum_{G'}\sigma_{-G'}(k+G')T_{G'}&=i\omega \rho_{0}T_{0}+i\sum_{G'}\rho_{-G'}\left(\omega+v_0G'\right)T_{G'}\\
-(k+G)\sigma_{G}kT_{0}-\sum_{G'}(k+G)\sigma_{G-G'}(k+G')T_{G'}&=i\omega \rho_{G}T_{0}+i\sum_{G'}\rho_{G-G'}\left(\omega+iv_0G'\right)T_{G'}
}{eq:sistGGp}
and we solve for $T_G$ from the second one,
\beq
T_{G'}=-\sum_{G}\chi_{G'G}\left[(k+G)\sigma_{G}k+i \omega \rho_{G}\right]T_0
\eeq
with
\beq
\chi_{GG'}=\left[(k+G)\sigma_{G-G'}(k+G')+i\left(\omega+v_0G'\right)\rho_{G-G'}\right]^{-1}
\eeq
and we introduce it in the first one, to obtain
\beq
\left(k^2\sigma_0+i\omega \rho_{0}-\sum_{G',G}\left[k\sigma_{-G'}(k+G')+i\left(\omega+v_0G'\right)\rho_{-G'}\right]\chi_{G'G}\left[(k+G)\sigma_{G}k+i \omega \rho_{G}\right]\right)T_0=0\\
\eeq
which can be expressed as
\beq
\left(k^2\sigma^*+i\omega \rho^*-i k C-i\omega k (S+S')\right)T_0=0
\eeq
with the effective parameters defined as
\subeqs{
\sigma^*(\omega,k)&=\sigma_0-\sum_{G',G}\sigma_{-G'}(k+G')\chi_{G'G}(k+G)\sigma_{G}\\
\rho^*(\omega,k)&=\rho_0-i\omega \sum_{G',G}\rho_{-G'}\chi_{G'G}\rho_{G}-iv_0 \sum_{G',G}G'\rho_{-G'}\chi_{G'G}\rho_{G}\\
S(\omega,k)&=\sum_{G',G}\sigma_{-G'}(k+G')\chi_{G'G}\rho_{G}\\
S'(\omega,k)&=\sum_{G',G}\rho_{-G'}\chi_{G'G}\sigma_{G}(k+G)\\
C(\omega,k)&=v_0\sum_{G',G}\rho_{-G'}G'\chi_{G'G}\sigma_{G}(k+G)
}{eq:effparams}

since $T_0$ is the average temperature $\langle T\rangle$, and replacing $k\to i\partial_x$ and $\omega\to -i\partial_t$, we can propose that the wave equation for the macroscopic temperature is
\beq
\sigma^*\frac{\partial^2 \langle T\rangle }{\partial x^2}= \rho^*\frac{\partial\langle T\rangle }{\partial t}+C\frac{\partial\langle T\rangle }{\partial x}-i(S+S')\frac{\partial^2 \langle T\rangle }{\partial x\partial t}
\eeq

The simpler case of modulation is a simple cosinus perturbation of the form
\subeqs{
\sigma(x-v_0t)&=\sigma_0\left[1+\Delta_\sigma\cos\frac{2\pi}{d}(x-v_0t)\right]\\
\rho(x-v_0t)&=\rho_0\left[1+\Delta_\rho\cos\frac{2\pi}{d}(x-v_0t)\right]
}{eq:harmonic}
where the mass density and conductivity changes periodically from $\rho_b=\rho_0(1-\Delta_\rho)$ to $\rho_a=\rho_0(1+\Delta_\rho)$ and from $\sigma_b=\sigma_0(1-\Delta_\sigma)$ to $\sigma_a=\sigma_0(1+\Delta_\sigma)$, respectively. In this case we have therefore only one Fourier component different than $0$, so that the $\chi_{GG'}$ matrix is diagonal with elements $\chi_{\pm}$ given by
\beq
\chi_\pm=\frac{d}{2\pi}\frac{d}{2\pi\sigma_0 \pm i v_0d\rho_0}
\eeq
Then it is easy to see that
\subeqs{
\sigma^*&=\sigma_0\left[1-\frac{8\pi^2\sigma_1^2}{4\pi^2\sigma_0^2+v_0^2d^2\rho_0^2}\right]\\
\rho^*&=\rho_0\left[1+\frac{2v_0^2d^2\rho_1^2}{4\pi^2\sigma_0^2+v_0^2d^2\rho_0^2}\right]\\
S&=S'=\frac{2iv_0d^2\rho_0\rho_1\sigma_1}{4\pi^2\sigma_0^2+v_0^2d^2\rho_0^2}\\
C&=v_0\frac{8\pi^2\sigma_0\sigma_1\rho_1}{4\pi^2\sigma_0^2+v_0^2d^2\rho_0^2}
}{eq:appeffective}
or
\subeqs{
\sigma^*&=\sigma_0\left[1-\frac{1}{2}\frac{\Delta_\sigma^2}{1+\Gamma^2}\right]\\
\rho^*&=\rho_0\left[1+\frac{\Gamma^2}{2}\frac{ \Delta_\rho^2}{1+\Gamma^2}\right]\\
S&=S'=\frac{\rho_0d}{2\pi}\frac{\Delta_\rho\Delta_\sigma}{2}\frac{i\Gamma}{1+\Gamma^2}\\
C&=\frac{2\pi\sigma_0}{d} \frac{\Delta_\rho\Delta_\sigma}{2}\frac{\Gamma}{1+\Gamma^2}
}{eq:appeffective2}
\end{widetext}

\end{document}